\documentclass{emulateapj}

\newcommand{\Swift}{{\it Swift}}
\newcommand{\NuSTAR}{{\it NuSTAR}}
\newcommand{\Suzaku}{{\it Suzaku}}
\newcommand{\XMM}{{\it XMM-Newton}}
\newcommand{\Chandra}{{\it Chandra}}

\usepackage{graphicx}
\usepackage{bm}
\usepackage{color}

\begin{document}

\title{Shedding Light on the Compton-thick Active Galactic Nucleus in the
Ultra-luminous Infrared Galaxy UGC 5101 with Broadband X-ray
Spectroscopy}

\author{Saeko Oda\altaffilmark{1},
Atsushi Tanimoto\altaffilmark{1},
Yoshihiro Ueda\altaffilmark{1},
Masatoshi Imanishi\altaffilmark{2,3,4},
Yuichi Terashima\altaffilmark{5},
Claudio Ricci\altaffilmark{6,7}}

\affil{
\altaffilmark{1}{Department of Astronomy, Kyoto University, Kyoto 606-8502, Japan\\}
\altaffilmark{2}{Subaru Telescope, 650 North A'ohoku Place, Hilo, HI 96720, USA\\}
\altaffilmark{3}{National Astronomical Observatory of Japan, 2-21-1 Osawa, Mitaka, Tokyo 181-8588, Japan\\}
\altaffilmark{4}{Department of Astronomy, School of Science, Graduate University for Advanced Studies (SOKENDAI), Mitaka, Tokyo 181-8588, Japan\\}
\altaffilmark{5}{Department of Physics, Ehime University, Bunkyo-cho, Matsuyama, Ehime 790-8577, Japan\\}
\altaffilmark{6}{Instituto de Astrofisica, Pontificia Universidad Catolica de Chile, Casilla 306, Santiago 22, Chile\\}
\altaffilmark{7}{Kavli Institute for Astronomy and Astrophysics, Peking University, Beijing 100871, China}}

\begin{abstract}

We report the broadband X-ray spectra of the ultra-luminous infrared
galaxy (ULIRG) UGC~5101 in the 0.25--100
keV band observed with \Swift/Burst Alert Telescope (BAT), \NuSTAR, \Suzaku, \XMM, and \Chandra. 
A Compton-thick AGN obscured with a hydrogen column density of
$\approx 1.3\times10^{24}$ cm$^{-2}$ is detected above 10 keV.
A spectral fit with a numerical torus model favors a large half opening
angle of the torus, $>41$ degrees, 
suggesting that the covering fraction of 
material heavily obscuring the X-ray source is moderate.
The intrinsic 2--10 keV luminosity
is determined to be $\approx 1.4\times 10^{43}$ erg s$^{-1}$, which is
$\approx$2.5 times larger than the previous estimate using only data below
10 keV with a simple spectral model.
We find that UGC~5101 shows
the ratio between the [O IV] 26 $\mu$m line and 2--10 keV luminosities
similar to those of normal Seyfert galaxies, along with other ULIRGs
observed with \NuSTAR, 
indicating that a significant portion of
local ULIRGs are not
really ``X-ray faint'' with respect to the flux of forbidden lines
originating from the narrow line region (NLR).
We propose a possible scenario that 
(1) the AGN in UGC 5101 is surrounded not only by 
Compton-thick matter located close to the equatorial plane
but also by Compton-thin ($N_\mathrm{H} \sim 10^{21}$ cm$^{-2}$) matter in
 the torus-hole region and (2) it is accreting at a high Eddington rate with a steep UV to X-ray
spectral energy distribution. Nevertheless, we argue that AGNs in many
ULIRGs do not look extraordinary (i.e., extremely X-ray faint), as
suggested by recent works, compared with normal Seyferts.

\end{abstract}

\keywords{galaxies: active -- galaxies: individual (UGC~5101) -- X-rays: galaxies}

\section{Introduction}

Ultra-luminous infrared galaxies (ULIRGs) radiate the bulk of their large
luminosities (L $>$ 10$^{12}$L$_{\odot}$) as dust emission. This
indicates that powerful energy sources, starbursts and/or active
galactic nuclei (AGNs), are hidden behind dust. ULIRGs are dominant
contributors to the cosmic infrared radiation density at $z>1.5$
(\citealt{cap07}, \citealt{got10}, \citealt{mur11}). Thus, distinguishing 
the energy sources in ULIRGs is critical to unveil
the history of star-formation and SMBH mass growth in the dust-obscured
galaxy population of the early universe. ULIRGs are major mergers of
gas-rich galaxies and have large amounts of concentrated molecular gas
and dust in their nuclei \citep{san96}. Hard X-ray ($E>10$
keV) observations (e.g., \citealt{teng09}, \citealt{teng15}) are very useful to
unambiguously detect hidden, even Compton-thick (with a hydrogen column
density of N$_{\rm H}$ $>$ 10$^{24}$ cm$^{-2}$) AGNs in ULIRGs thanks to
their strong penetrating power, unless it largely exceeds $N_{\rm H}
\sim 10^{25}$ cm$^{-2}$.

\begin{deluxetable*}{ccccccc}
\tablecaption{Detailed information of UGC 5101 \label{info}}
\tablewidth{0pt}
\tablehead{Object & Redshift   &  $f_{\rm 12}$  &  $f_{\rm 25}$  &  $f_{\rm 60}$  &  $f_{\rm 100}$  &  log $L_{\rm IR}$ (log $L_{\rm IR}$/$L_{\odot}$)  \\ 
 &  & [Jy] &  [Jy] & [Jy] & [Jy] & [ergs s$^{-1}$] \\
(1) & (2) & (3) & (4) & (5) & (6) & (7)\\
}
\startdata
UGC 5101 & 0.040 & 0.25 & 1.03 & 11.54 & 20.23 & 45.5 (12.0) 
\enddata
\tablecomments{
(1) Object. 
(2) Redshift.
(3)--(6)  f$_{12}$, f$_{25}$, f$_{60}$, and f$_{100}$ are the {\it IRAS FSC} fluxes at 12, 25, 60, and 100 $\mu$m, respectively.
(7) Decimal logarithm of the infrared (8$-$1000 $\mu$m) luminosity in ergs s$^{-1}$ calculated as follows:
$L_{\rm IR} = 2.1 \times 10^{39} \times$ D(Mpc)$^{2}$ $\times$ (13.48 $\times$ $f_{12}$ + 5.16 $\times$ $f_{25}$ + $2.58 \times f_{60} + f_{100}$) ergs s$^{-1}$ \citep{san96}.
The values in parentheses are the decimal logarithms of the infrared luminosities in units of solar luminosities. 
}
\end{deluxetable*}

UGC 5101 ($z=$0.040), the subject of this paper, is a well-studied
nearby ULIRG
with an infrared (8--1000 $\mu$m) luminosity of L$_{\rm IR}$ $\simeq$
10$^{12}$L$_{\odot}$ (Table \ref{info}).
It is optically classified as a LINER based on systematic investigation
of luminous infrared galaxies \citep{vei95}, but there are some
arguments that obscured AGN signatures (i.e., Seyfert 2) are detected in
optical spectra using some combination of optical emission lines
\citep{gon99,yua10}. \citet{arm04} detected high-excitation forbidden
emission lines in the mid-infrared 5--30 $\mu$m, such as [Ne V] 14.3
$\mu$m and [O IV] 26 $\mu$m, which are naturally explained by the
so-called narrow line regions (NLRs) photoionized by an AGN
\citep{rob96}.  Signatures of an obscured AGN in UGC~5101 have been also
seen in various methods, which do not rely on the emission from the
NLRs, at the wavelengths of low dust extinction. A compact high
surface-brightness emission was detected in the near-infrared 1--2 $\mu$m
\citep{sco00} and mid-infrared 12.5 $\mu$m \citep{soi00}, which may
indicate the presence of a luminous AGN. Infrared 2.5--5 $\mu$m
\citep{idm01,imm03,ima08} and 5--35 $\mu$m \citep{arm04,nar09,vei09}
spectral shapes, including the broad polycyclic aromatic hydrocarbon (PAH) emission features, are
naturally explained by a luminous obscured AGN that contributes
importantly to the observed flux at 2.5--35 $\mu$m. Millimeter
interferometric observations using the Nobeyama Millimeter Array (NMA)
showed an enhanced HCN J=1--0 to HCO$^{+}$ J=1--0 emission-line flux
ratio, compared to starburst-dominated galaxies \citep{ink06}, which is
often seen in galaxies containing energetically important AGNs
\citep{koh05,kri08}. Radio VLBI very high-spatial-resolution
observations at 18 cm also suggest the presence of a luminous AGN
\citep{smi98,lon03}.

The AGN contribution to the bolometric luminosity in UGC~5101 is yet
uncertain, however.
According to \citet{arm07}, 
the estimated AGN contribution
based on the mid-infrared
NLR-originated forbidden emission lines is less than 10\%, 
which is substantially
smaller than those 
by 
other energy diagnostic methods using mid-infrared
continuum slope and equivalent width of the PAH emission features,
$>50$\%.
Because these mid-infrared lines emitted from the NLR are expected
to be little affected by extinction, it is unlikely that weak NLR lines
are attributed to dust absorption by interstellar gas in the host galaxy.
Instead, a possible hypothesis is that 
the AGN in UGC 5101 is almost fully surrounded by dust located on
inner scale with respect to the NLR, i.e., UGC 5101 contains a so called
``buried AGN''.
In a buried AGN, the NLR is under-developed  compared to the majority
of optical Seyfert galaxies because a large fraction of ionizing UV
photons from the central AGN may be absorbed by dust at the very inner
part before reaching the NLR.
On the basis of 2--10 keV X-ray observations using
\XMM \ and \Chandra, \citet{ima03} detected X-ray emission that is
likely to originate
in a Compton-thick AGN. 
They estimated the intrinsic (absorption-corrected) 2--10 keV
luminosity to be $\approx 6\times10^{42}$ erg s$^{-1}$.
\footnote{This luminosity has been converted from the original value
$\approx 5\times10^{42}$ erg s$^{-1}$ obtained 
for a Hubble constant of $H_0 = 75\ \mathrm{km\
s^{-1}\ Mpc^{-1}}$ to that for $H_0 = 70\ \mathrm{km\ s^{-1}\ Mpc^{-1}}$.
}
However, due to the limitation of the X-ray spectral coverage at $<$10 keV,
large ambiguity remains about it.
This source is also detected in the \Swift/Burst Alert Telescope (BAT)
70-month survey covering the 14--195 keV band, and is listed as a
Compton-thick AGN candidate (\citealt{ric15}).

In this paper, we report the broadband X-ray spectra of UGC~5101
covering the 0.25--100 keV band, observed with \Swift/BAT, \NuSTAR, 
\Suzaku, \XMM, and \Chandra. A combination of all available data
provides us with an ideal opportunity to scrutinize the properties of
such a Compton thick AGN in UGC 5101 in a much more reliable manner than
in previous studies.
We organize this paper as follows. Section 2 presents the details of the
observations and reduction. Section 3 describes the results of spectral
analysis. Section 4 discusses implications from our results. Section 5
states the conclusion of this paper.  We adopt $H_0 = 70\ \mathrm{km\
s^{-1}\ Mpc^{-1}}\ ,\ \Omega_\mathrm{m} = 0.27\ ,\ \Omega_{\lambda} =
0.73$. The solar abundances by \citet{and89} and the photoelectric
absorption cross-sections by \citet{bal92} are assumed throughout this paper. All
errors attached to spectral parameters correspond to 90\% confidence
limits for a single parameter of interest.

\tabletypesize{\scriptsize}
\begin{deluxetable*}{cccccc}
\tablecaption{Summary of observations \label{obs}}
\tablewidth{0pt}
\tablehead{  & Instrument  &  Observation ID  &  Start Time$^*$  &  End Time$^*$ &  Exposure$^*$  \\
  &  &  & [UT] & [UT] & [ks] \\
} 
\startdata
\Swift  &  BAT  &  SWIFT J0935.9+6120 &  2004 December 16 00:00  & 2010 September 30 23:59  &  13428.3 \\
 \NuSTAR  &   FPMA, FPMB  &  60001068002  &  2014 March 21 22:26   &  2014 March 22 09:01 & 18.3  \\
\Suzaku   &  XIS-0,1,2,3, HXD  &  701002010  &   2006 March 31 04:18  &  2006 April 1 08:25  &   49.0  \\
   &   & 701002020  & 2006 October 31 18:19  &  2006 November 1 09:30  &  42.2  \\
\XMM  &  EPIC-MOS1,2, EPIC-pn  & 0085640201  &  2001 November 12 09:02  & 2001 November 12 18:35  &  33.9  \\
\Chandra  &  ACIS  &  2033  &  2001 May 28 07:02  &  2001 May 28 21:21  &  42.1 
\enddata
\tablenotetext{*}{
Based on the information of 
FPMA for \NuSTAR, XIS-0 for \Suzaku, and EPIC-MOS1 for \XMM. 
}
\end{deluxetable*}

\tabletypesize{\footnotesize}
\begin{deluxetable*}{cccccc}
\tablecaption{Summary of data reduction  \label{reduction}}
\tablewidth{0pt}
\tablehead{
 & Instrument & source radius & background radius & encircled energy fraction & net count rate  \\
  &  & [arcsec] & [arcsec] & [\%] & [10$^{-3}$ count s$^{-1}$] \\
 & & (1) & (2) & (3) & (4) }
\startdata
\Swift & BAT & -- & -- & -- & 0.01 \\
\NuSTAR &  FPMs & 50 & 50 & 68 &  8.51 \\
\Suzaku & FI-XISs & 60 & 100 & $\sim$50 & 2.84 \\
  & BI-XIS & 60 & 100 & 44 & 2.16 \\
\XMM & EPIC-MOSs & 22.5 & 60 & $\sim$80 & 6.94 \\
 & EPIC-pn & 22.5 & 60 & 84 & 26.8 \\
\Chandra & ACIS & 8.8  &  9.8--24.6  & 96 &  8.71 
\enddata
\tablecomments{
(1) The radius of the extraction region for the source spectrum.
 (2) The radius (or inner and outer radii) of the extraction
 region for the background spectrum.
 (3) The encircled energy fractions at 4.5 keV for \Suzaku, at 6 keV for \XMM, and at 6 keV for \Chandra.
 (4) The net count rate after background subtraction.
 }
\end{deluxetable*}

\begin{figure}
\begin{center}
\plotone{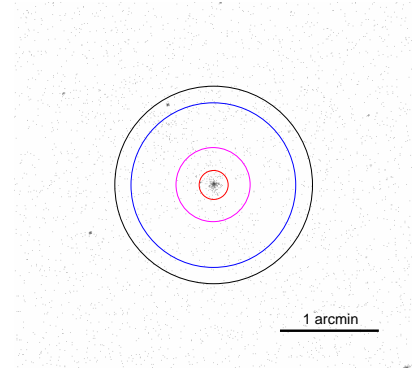}
\end{center}
\caption{Spectral extraction regions marked on the 
 \Chandra \  image of UGC 5101 in the 0.3--7 keV band. The image size is 4 arcmin square.
 The red, magenta, blue, and black circular regions
 correspond to radii of 8.8 (\Chandra), 22.5 (\XMM), 50 (\NuSTAR), and 60
arcsec (\Suzaku), respectively.}
\label{extract}
\end{figure}

\section{Observations and Data Reduction}

We utilize the data observed with \NuSTAR, \Suzaku,
\XMM, and \Chandra. 
To better constrain the broadband spectrum, we also use
the time-averaged spectrum obtained in the \Swift/BAT 70-month survey
(http://swift.gsfc.nasa.\\gov/results/bs70mon). 
Table~\ref{obs} lists the details of the observations. 
The spectral extraction regions and net count rate 
for each observatory are summarized in Table~\ref{reduction}.
These are chosen to maximize the signal-to-noise (S/N) ratio
by considering the size of the point spread function, and to 
minimize the contamination from other sources.
Figure~\ref{extract} displays the \Chandra \ image in the 0.3--7 keV band
centered at UGC 5101 (see Section~2.4 for detail), where the source extraction
regions are overlaid.

\subsection{NuSTAR}

The Nuclear Spectroscopic Telescope Array (\NuSTAR) mission,
launched on 2012 June 13, is the first focusing high-energy X-ray
telescope, 
which covers the energy range of 3--78 keV \citep{har13}.
It carries two co-aligned hard X-ray
telescopes, each of which has its own focal plane module (FPMA and
FPMB, FPMs). UGC 5101 was observed with \NuSTAR \ on 2014 March 21 and
22 for a net exposure of 18.3 ksec. We reduce the data, using HEAsoft
version 6.17 and calibration database version 2015 September 4. We use
the scripts \textit{nupipeline} and \textit{nuproducts} to make the
spectra of FPMA and FPMB. Photon events in a circular region with a
radius of 50 arcsec around the target are selected. The
background spectrum is extracted from a nearby source-free circular region with the
same size as the source region. 
We have verified that the FPMA and FPMB spectra give
fully consistent results including the normalizations. Then, 
to improve the photon statistics in a single spectrum, 
the data of FPMA and FPMB are combined by using the script \textit{addascaspec}.

\begin{figure}
\begin{center}
\plotone{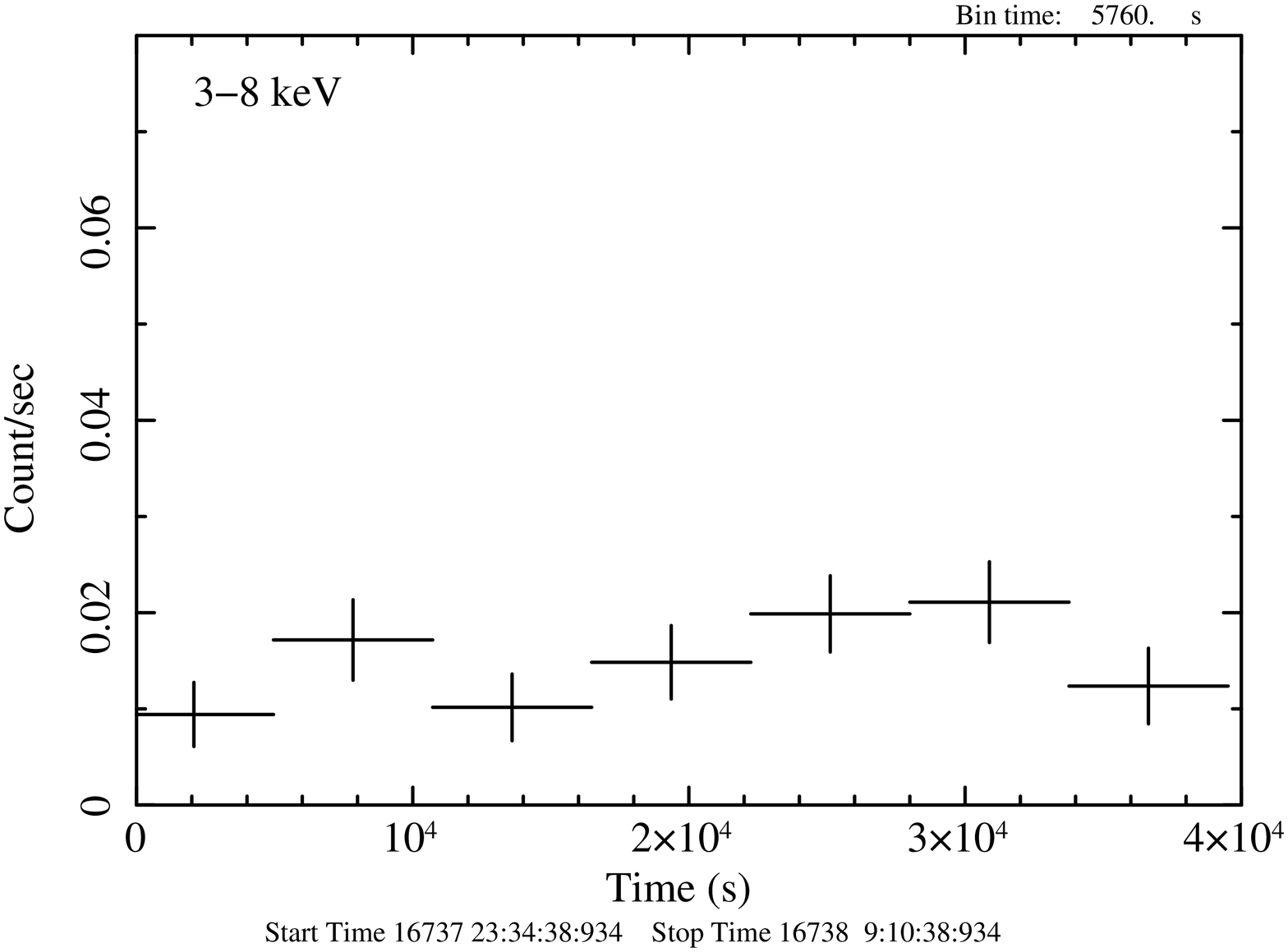}
\plotone{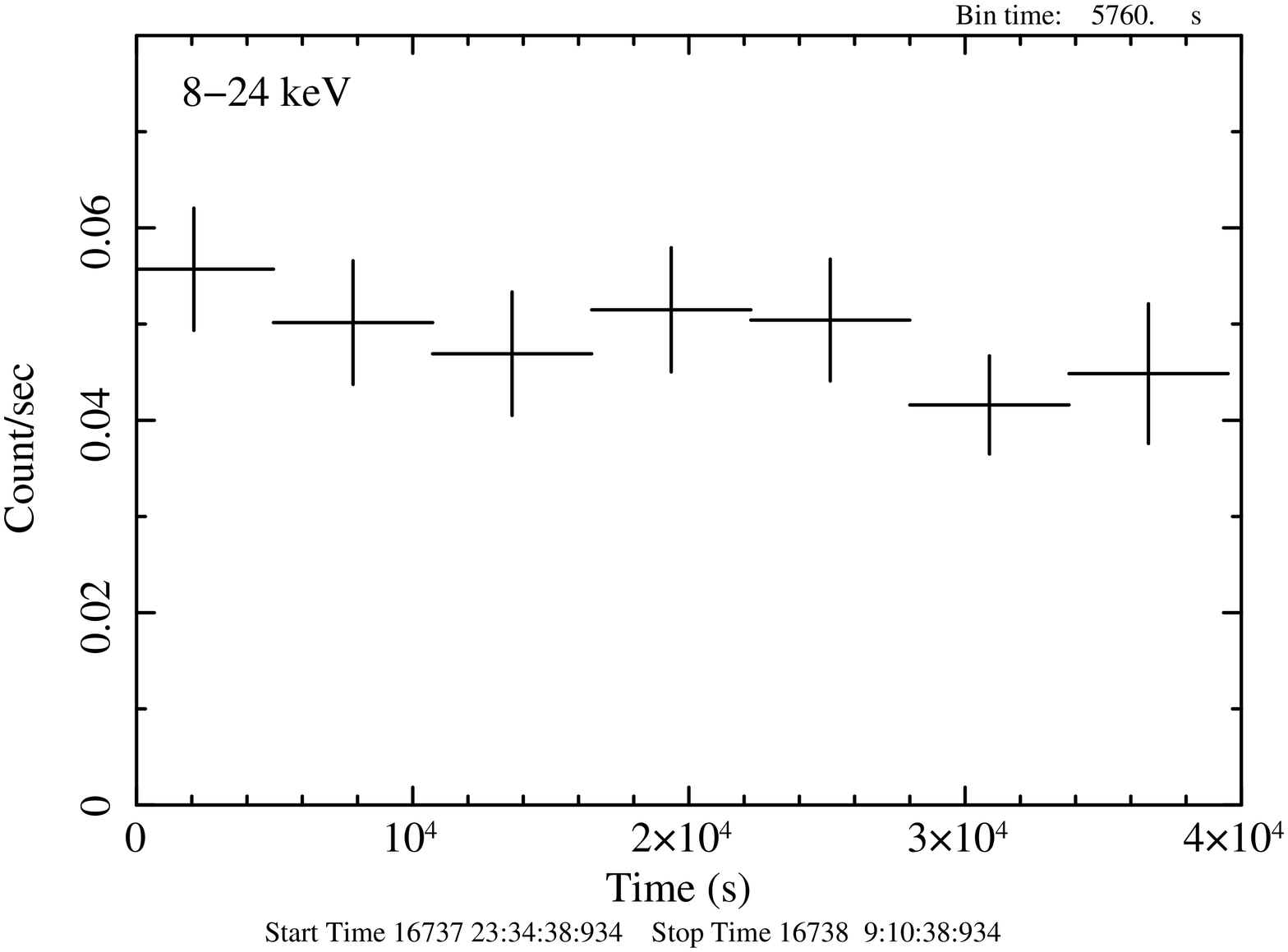}
\end{center}
\caption{\NuSTAR \ light curves in the 3--8 keV and
 8--24 keV bands. The bin size is 5760 sec.}
\label{lc}
\end{figure}

Figure~\ref{lc} plots the background-subtracted \NuSTAR\ light curves in
the 3--8 and 8--24 keV bands. The bin size
is set to be 5760 sec (96 minutes), the orbital period of the satellite,
to avoid any possible modulation related to orbital phase.
The source and
background regions are the same as those used for the spectral
extraction. 
The FPMA and FPMB light curves are summed by
using the \textit{lcmath} task.
Performing a $\chi^2$ test to each light curve with a constant model, 
we find that time variability during the \NuSTAR\ observation was
not significant at $>90\%$ confidence limits on timescales
longer than 96 minutes in both energy bands.

\subsection{Suzaku}

\Suzaku \ is the fifth observatory in a series of Japanese X-ray
astronomy satellites \citep{mit07}. It carries four X-ray CCD cameras
called the X-ray Imaging Spectrometers (three Front-side-Illuminated
XISs or FI-XISs, XIS-0, XIS-2, and XIS-3, and one Back-side-Illuminated
XIS or BI-XIS, XIS-1, \citealt{koya07}) and a non-imaging instrument that consist of Si
PIN photo-diodes and Gadolinium Silicon Oxide (GSO) scintillation
counters, called the Hard X-ray Detector (HXD, \citealt{tak07}). 
UGC 5101 was observed
with \Suzaku \ on 2006 March 31, April 1 and 2006 October 31, November 1 for exposures of 49.0
ksec and 42.2 ksec, respectively.  We analyze the data in a standard
manner, using HEAsoft version 6.17 and calibration database released on
2015 October 5.

\subsubsection{XIS}

The XISs cover the energy range below $\approx$10 keV. To extract the
spectra, we accumulate photon events in the circle centered at the
source with a radius of 60 arcsec. The background is taken from a
source-free region in the field of view. In the source extraction
region, the \Chandra \ image (Figure~\ref{extract}) shows another, much fainter
point-like source, whose distance from the target is
$\sim 55$ arcsec, and therefore we add its estimated contribution to
the background by analyzing the \Chandra \ data.  We confirm that the
contribution of this source is negligible in the spectra of the other
instruments, including \NuSTAR/FPMs.

We sum up the spectra of XIS-0, XIS-2 and XIS-3, whereas the XIS-1
spectrum was treated independently. Since no significant variability in
the spectra was found between the two observation epochs, we make the
time averaged spectra by combining them.

\subsubsection{HXD}

We also examine the data of HXD/PIN,
which covers the energy range of 14--70 keV and hence are suitable for detecting the
transmitted component from a Compton-thick AGN.
The ``tuned'' background file (version 2.0) released
from the \Suzaku \ team is used as the non Xray background (NXB), while the
Cosmic X-ray Background (CXB) is estimated by assuming the spectral form
presented in \citet{bol87}. The response files
ae\_hxd\_pinhxnome3\_20080129.rsp and ae\_hxd\_pinflate3\_20080129.rsp
are used to estimate the source flux and the CXB, respectively.

No significant signals from UGC 5101 are detected with HXD/PIN. The
count rate in the 15--40 keV after subtraction of the NXB and CXB is
0.00085 counts s$^{-1}$, which is 0.30\% of that of the NXB. The
3$\sigma$ upper limit on the count rates is 0.0085 counts s$^{-1}$,
which is 3\% of the NXB rate \citep{fuk09}.  This upper limit
corresponds to a hard X-ray flux of $3.8 \times 10^{-12}$ erg cm$^{-2}$
s$^{-1}$ in the 15--40 keV band for Model Ib (see Section~3.1.2),
which is 1.8 and 1.2 times larger than the actual fluxes measured with
\Swift/BAT and \NuSTAR, respectively.

\subsection{XMM-Newton}

We analyze the \XMM \ EPIC data obtained on 2001 November 12.
To take into account the latest calibration, the data are reprocessed and analyzed
by using the Science Analysis System (SAS) version 14.0.0 and the latest
calibration files (CCF) as of 2015 October.  We examine light curves of
PATTERN=0 events in the energy range of 10--12 keV (for EPIC-pn) or
above 10 keV (for EPIC-MOS), and find that the background fluxes are
relatively stable, although some flux increases during the observation
are seen. We apply no filter to discard such time intervals. The source
and background spectra are extracted from a circular region with a
radius of 22.5 arcsec and a source free region in the same CCD chip,
respectively. Only PATTERN $<$= 4 and PATTERN $<$=12 events are used for
EPIC-pn and EPIC-MOS, respectively. The redistribution matrix function
(RMF) and auxiliary response file (ARF) are made using the SAS tasks
rmfgen and arfgen.

\subsection{Chandra}

A \Chandra \ observation of UGC 5101 was performed on 2001 May 28.  The
ACIS was operated in the FAINT mode. The \Chandra \ data are reprocessed
and analyzed using the \Chandra \ Interactive Analysis of Observations
(CIAO) software package version 4.7 combined with the calibration CALDB
version 4.6.9. 
UGC 5101 is located in the ACIS-S3 back-illuminated CCD
chip. 
The data are screened by applying the lc\_clean script for the
light curve in the 0.5--7 keV band of the whole ACIS-S3 excluding bright
sources. Only the events with grade 0, 2, 3, 4, 6 are used in the
following analysis. The source spectrum is extracted from a circular
region with a radius of 8.8 arcsec centered at the position of UGC
5101. 
We find from the 0.3--7 keV image (Figure~\ref{extract}) that there is 
diffuse emission extending $\sim 18$ arcsec
in the north-south direction and $\sim 8$ arcsec in the east-west
direction around the point-like source.
Most of these photon counts are also included in the spectral
extraction region.
The background spectrum is taken from an annular region around
the source. The spectra, RMF and ARF are created by using the script
\textit{specextract} in the CIAO package.
Benefiting from its unsurpassed high image quality (full
width at half maximum $\sim 0.5$ arcsec), \Chandra \ data are suitable for
studying possible contamination from other sources (see Section 2.2.1).

\begin{figure}
\begin{center}
\plotone{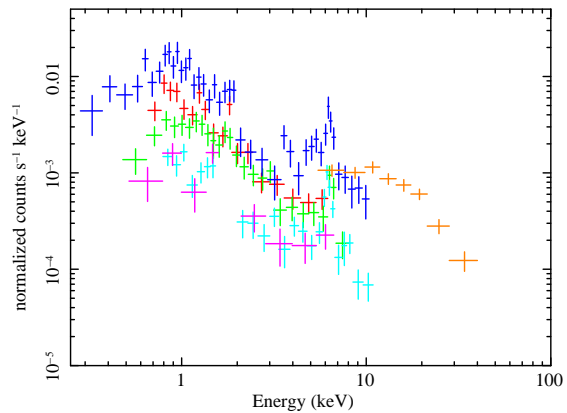}
\end{center}
\caption{Observed spectra of UGC 5101 folded with the energy responses.
The orange, light blue, magenta, green, blue and red crosses
represent the data of FPMs, FI-XISs, BI-XIS, EPIC-MOS, EPIC-pn, and
ACIS, respectively. }
\label{rawdata}
\end{figure}

\section{Spectral Analysis}

We simultaneously analyze the set of
seven spectra (\Swift/BAT, \NuSTAR/FPMs, \Suzaku/FI-XISs and BI-XIS, \XMM/EPIC-pn
and EPIC-MOSs, and \Chandra/ACIS). Figure~\ref{rawdata} shows the observed
spectra folded with the energy responses 
except for \Swift/BAT.
A prominent Fe K$\alpha$ line at 6.4~keV can be clearly seen in Figure~\ref{rawdata}.

For spectral analysis, we use the data of BAT, FPMs, FI-XISs, BI-XIS,
MOSs, pn, and ACIS
in the energy bands of 14--100 keV, 5.5--40 keV, 0.8--11 keV, 0.5--7 keV, 0.5--8 keV,
0.25--10 keV, and 0.65--6 keV, respectively, where 
the signal-to-noise ratio is sufficiently high.
The data of FI-XISs and BI-XIS in the 
1.6--1.9 keV band are excluded to avoid calibration uncertainties.
The spectral bins
are merged to contain at least 100, 50, 20, and 20 counts for each
spectrum of \NuSTAR, \Suzaku, \XMM, and \Chandra,
respectively. Spectral fit is performed on XSPEC with the $\chi^2$
minimization method. To take into account the Galactic absorption,
\textbf{phabs} is always multiplied to the spectral model, whose
hydrogen column density is fixed at the value based on the
$\mathrm{H_{I}}$ map \citep{kal05}. We multiply a constant factor (\textbf{const0}; $N_\mathrm{FI}$, $N_\mathrm{BI}$, $N_\mathrm{MOS}$, $N_\mathrm{pn}$, and $N_\mathrm{ACIS}$ for FI-XISs, BI-XIS, EPIC-MOSs, EPIC-pn, and ACIS, respectively) to
each spectrum in order to correct for cross calibration uncertainties
in the absolute flux among different instruments.
It is fixed at unity
for the BAT spectrum,
which is adopted as calibration references in
this paper. 
For the other instruments, it is allowed to vary within a reasonable range, 0.8--1.2.
Furthermore, we link the \textbf{const0}
factor of FPMs to those of MOSs because their relative
normalizations are well calibrated within a few percents \citep{mad15}; 
we confirm that these few percent uncertainties have 
negligible effects on the spectral parameters. 
In addition, we also consider possible time variability 
of the AGN emission (see below) with respect to the averaged
flux measured with \Swift/BAT over 70 months.

\subsection{Analytical Models}

\subsubsection{Model Ia}

First we start with Model B in \citet{ima03} adopted to fit the
\XMM \ and \Chandra \ spectra. The model consists of a primary, 
heavily absorbed power law with an exponential cutoff (hereafter ``cutoff power law''), 
a less absorbed cutoff power law (or a
scattered component),
a narrow gaussian representing an iron-K emission line, and
an optically-thin thermal component.
In XSPEC, this model (hereafter ``Model Ia'') is described as:
\begin{eqnarray}
&& \mathbf{const0} * \mathbf{phabs}* \{ \mathbf{zphabs1}     \nonumber     \\  
&*& ( \mathbf{const1} * \mathbf{zphabs2} * \mathbf{zpowerlw} *\mathbf{zhighect}   \nonumber \\ 
&+& \mathbf{const1}* \mathbf{zgauss}  \nonumber \\
&+& \mathbf{const2} * \mathbf{zpowerlw} * \mathbf{zhighect} ) \nonumber \\
&+&  \mathbf{apec}.  \} 
\end{eqnarray}
We fix a cutoff energy 
\footnote{This corresponds to an e-folding energy ($E_{\rm f}$) in the
\textbf{zhighect} model when $E_{\rm c}$ is set to be zero.} 
at 360 keV, the value adopted in the torus model by \citet{ike09} (Model II, see Section 3.2),
because it cannot be well constrained from the data; 
variation of the cutoff energy within a reasonable
range (100--500 keV) has little influence on the spectral parameters in
Models~Ia and Ib (Section~3.1.2).
The constant factor \textbf{const1} is introduced to
account for possible time variability of the AGN component
relative to the flux measured with \Swift/BAT. Except for \NuSTAR, 
it is found to be consistent with unity 
within the expected instrumental cross normalizations (\textbf{const0}).
Hence, it is left free for the \NuSTAR \ spectrum
($N_\mathrm{FPM}^\mathrm{time}$) and is fixed at unity
for the other spectra in the final spectral fit.
The \textbf{const2} factor gives the scattering fraction, $f_\textrm{scat}$.
We define it as the ratio of the unabsorbed fluxes at 1 keV between the
scattered and primary power law components, whose
normalizations are tied together.
The photon index of the scattered component is set to be 
independent of that of the primary one.
The weak absorption
(\textbf{zphabs1}, $N_\mathrm{H}^\mathrm{host} \sim 1\times10^{21}$ cm$^{-2}$) to the second power-law component
is significantly required from the data, although this fact was
(mistakenly) not mentioned in \citet{ima03}; otherwise, the model
leaves large negative residuals below $\sim$1 keV.
We confirm that the presence of this absorption 
is robust against modelling of the scattered component
in both Models~Ia and Ib; it is still required when we include emission
lines and radiative recombination continua of a photoionized plasma whose
flux ratios are fixed to those in NGC 1068 (\citealt{gua07}), and when the
photon index is tied to that of the primary component.

This phenomenological model
gives an acceptable fit of our combined spectra ($\chi^2$/d.o.f. =
123.40/117).
Table~\ref{par1} lists the best fit parameters,
the observed fluxes and intrinsic (de-absorbed) luminosities in the
2--10 keV and 10--50 keV bands, and the equivalent width of the 
iron-K line with respect to the total continuum.
The fluxes and luminosities are based on 
the \Swift/BAT spectrum,
for which \textbf{const0} and \textbf{const1} are set unity.
Figure~\ref{spectra} and Figure~\ref{eem}
present the unfolded spectra and best-fit models for Model Ia, respectively.
The line center energy becomes 6.46 keV. It is consistent with a
fluorescence line from cold matter as the main component, although weak
contribution from ionized lines (e.g., \citealt{iwa11}) is possible.

\subsubsection{Model Ib}

Next, we consider a more physically motivated model than Model Ia.
Firstly, a Compton reflection continuum from the torus should be
present, which accounts for the origin of the narrow iron-K$\alpha$
line. Secondly, in the case of a Compton-thick torus, 
we should take into account 
Compton scattering for the transmitted component
in addition to photoelectric absorption.
Assuming an extreme case where the absorber is located only along the
line of sight (i.e., photons scattering into the line of sight from
other directions are ignorable), this can be modelled by multiplying
\textbf{cabs} to the transmitted component.
We thus adopt the following analytical model as our ``Model Ib''.
Similar spectra are often observed from ULIRGs (e.g., \citealt{teng15}). 
This model is expressed in the XSPEC terminology as: 
\begin{eqnarray}
&& \mathbf{const0} * \mathbf{phabs}* \{ \mathbf{zphabs1}     \nonumber     \\  
&*& ( \mathbf{const1} * \mathbf{zphabs2} * \mathbf{cabs} * \mathbf{zpowerlw} *\mathbf{zhighect}   \nonumber \\ 
&+& \mathbf{const1} * \mathbf{pexmon}   \nonumber \\
&+& \mathbf{const2} * \mathbf{zpowerlw} * \mathbf{zhighect} ) \nonumber \\
&+& \mathbf{apec}.  \}  
\end{eqnarray}
(1) The first term represents the transmitted component from the AGN,
    modelled by an absorbed cutoff power law with a cutoff energy of 360 keV.
    The \textbf{const1} factor, which accounts for time
    variation with respect to the \Swift/BAT spectrum, 
    is linked to that for the reflection (second)
    component.
(2) The second term approximately describes a reflection component from
    the torus.
    Here we utilize the \textbf{pexmon} code \citep{nan07}.
    This combines the \textbf{pexrav} model,
    which calculates a Compton-reflection continuum from optically
    thick, cold matter (\citealt{mag95}), with fluorescence lines from heavy elements.
    The relative intensity is defined as $R \equiv
    \Omega / 2 \pi$, where $\Omega$ is the solid angle of the
    reflector. We leave $R$ free within a range of $-2 \le R < 0$ and
    fix the inclination angle at 60 degree 
    as a representative value for a geometrically thick torus (e.g.,
    \citealt{egu09}; \citealt{kaw16}).
    The negative values of $R$ indicate that the \textbf{pexmon} model
    represents only the reflection component.
    The photon index, cutoff energy, and normalization
    are linked to those of the transmitted component.
   We do not apply absorption to the reflection component, which
   is not significantly ($>$90\% confidence level) required from the data.
    The same constant factor \textbf{const1} as the first term is multiplied to the second
    term in order to take into account time variability, 
    although this assumption may not be trivial, depending
    on the location of the reflector.
\footnote{
    We obtain $R=-0.11^{+0.07}_{-0.19}$ when the \textbf{const1} factor is not multiplied to
    the \textbf{pexmon} component. Thus, the conclusion of small $R$ is unchanged.
}
(3) The third term is the scattered component of the intrinsic AGN
    spectrum, modelled with a cutoff power-law whose 
    parameters except photon index are linked to those in the transmitted
    component, multiplied by \textbf{const2} ($f_\mathrm{scat}$).
    A weak
    absorption (\textbf{zphabs1}) is multiplied to the first, second and
    third terms.  We assume that the third and fourth components
    did not vary among the multiple observation epochs, considering the
    large spatial scales of the emitting regions.
(4) The fourth term accounts for optically-thin thermal emission in the
    host galaxy. Here we ignore any intrinsic absorption.

We find that this model also well reproduces the combined spectra
covering the 0.25--100 keV band ($\chi^2$/d.o.f = 130.59/118). Table \ref{par1} lists the best-fit
parameters, the observed fluxes and intrinsic luminosities in the 2--10 keV
and 10--50 keV bands based on the \Swift/BAT spectrum,
and the 6.4 keV line equivalent-width predicted
from the best-fit model.
The unfolded spectra and best-fit models are presented in Figures \ref{spectra}
and \ref{eem}, respectively. The line-of-sight column density of the
transmitted component is found to be $N_\mathrm{H}^\mathrm{LS} = 1.26^{+0.33}_{-0.31}
\times 10^{24}$ cm$^{-2}$, which 
identifies UGC 5101 as a Compton-thick AGN,
consistent with the result of
Model Ia.

\subsection{Torus Model}

As an even more realistic picture, we utilize the torus model by \citet{ike09}
(hereafter ``Ikeda torus model''), a numerical spectral model
based on Monte Carlo simulations. 
The Ikeda torus model assumes 
a nearly spherical torus consisting of neutral, cold material and has
two cone-shaped holes along the polar axis. The geometry is determined
by the torus half-opening angle ($\theta_\mathrm{op}$ within a range of
0--70 degree), inclination angle ($\theta_\mathrm{incl}$ within a range
of 1-89 degree), and the ratio of the inner to outer radii,
$r_\mathrm{in} / r_\mathrm{out}$, which is fixed at 0.01.  The variable
half-opening angle is one of the advantages of this model among similar
numerical models, because it is expected not to be constant in AGNs. The
incident spectrum is a cutoff power law with a cutoff energy of 360
keV. The other free parameters of this model are photon index
($\Gamma_{\rm AGN}$) and column density along the equatorial plane
($N_\mathrm{H}^\mathrm{Eq}$), which are variable between 1.5--2.5 and
between $10^{22}$--$10^{25}$ cm$^{-2}$, respectively (see \citealt{ike09} for
details). We fix the inclination angle at 80 degrees, in order to assure
a type-2 AGN geometry, while the half opening angle is left free. The
equatorial column density, photon index, and normalization are linked to
those of the transmitted component.

In the XSPEC terminology, the total model (Model~II) is represented as follows:
\begin{eqnarray}
&& \mathbf{const0} * \mathbf{phabs}* \{ \mathbf{zphabs1}     \nonumber     \\  
&*& ( \mathbf{const1} * \mathbf{torusabs} * \mathbf{zpowerlw} *\mathbf{zhighect}   \nonumber \\ 
&+& \mathbf{const1} * \mathbf{atable \{refl1 \_ torus.fits \}}  \nonumber \\
&+& \mathbf{const1} * \mathbf{atable \{refl2 \_ torus.fits \}} \nonumber \\
&+& \mathbf{const1} * \mathbf{atable \{refl \_ fe \_ torus.fits\} }    \nonumber \\
&+& \mathbf{const2} * \mathbf{zpowerlw} * \mathbf{zhighect} ) \nonumber \\
&+& \mathbf{apec}.  \} 
\end{eqnarray}
The model consists of six components: (1) the transmitted component
from the AGN, 
(2) the torus reflection component absorbed by the near-side torus, (3) 
that not absorbed by the near-side torus, (4) the iron-K$\alpha$ emission
line, 
(5) the scattered component (with a photon index $\Gamma_\mathrm{scat}$ and
a scattering fraction of $f_\mathrm{scat}$), 
and (6) the optically-thin thermal component. 
Similarly to the case of analytical models, we consider 
time variability of the (1)-(4) components for the \NuSTAR \  data.
We do not utilize the energy band below 0.7 keV with Model~II, 
where the Ikeda torus model is not available.
Hence, the absorption from the host galaxy and the parameters
of the optically-thin thermal component are fixed at
the same values as in Model Ib, which cannot be well constrained with the
data above 0.7 keV. We confirm that changing the fixed parameters within
the uncertainties does not affect the conclusions.

\begin{figure}
\begin{center}
\plotone{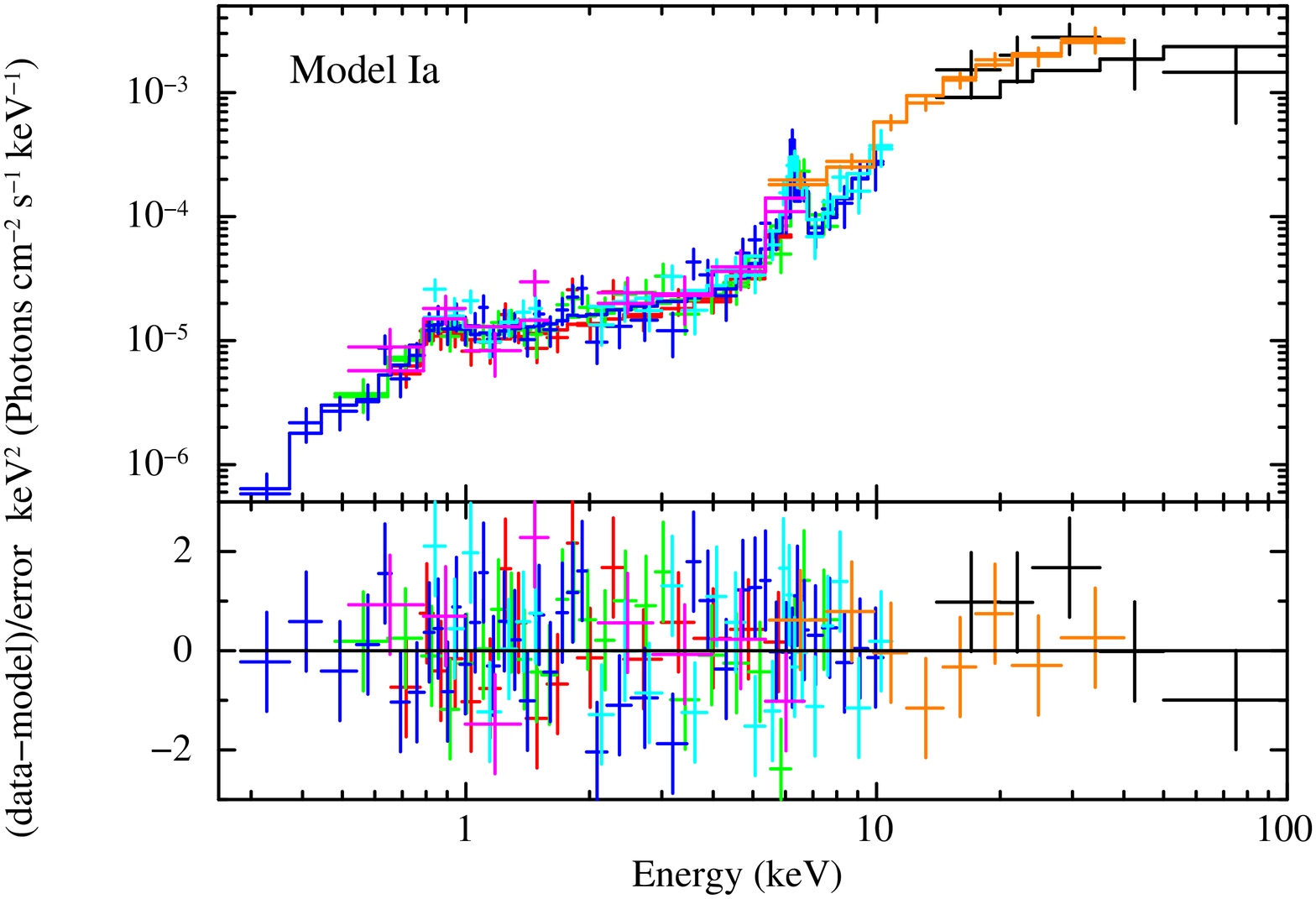}
\plotone{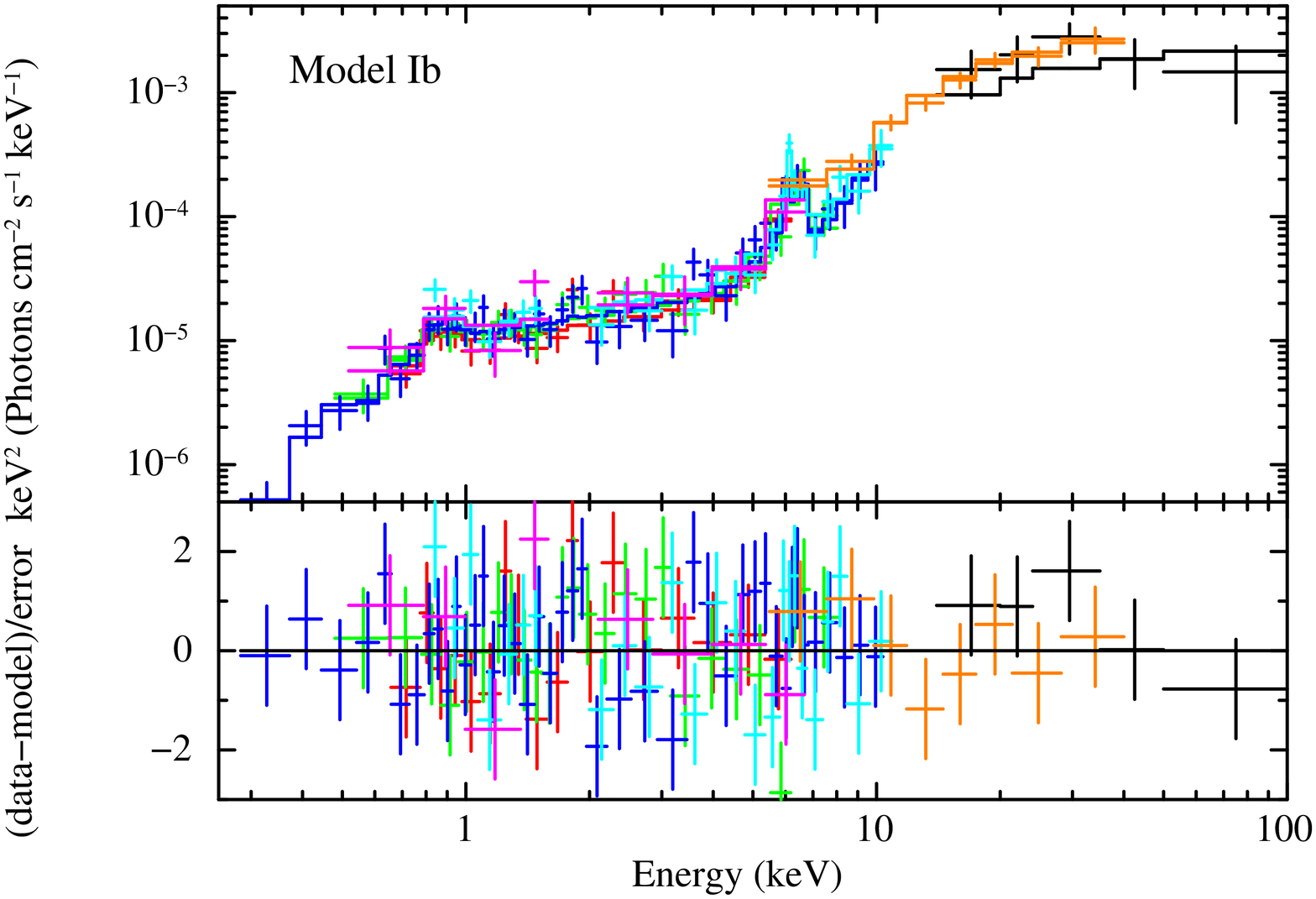}
\plotone{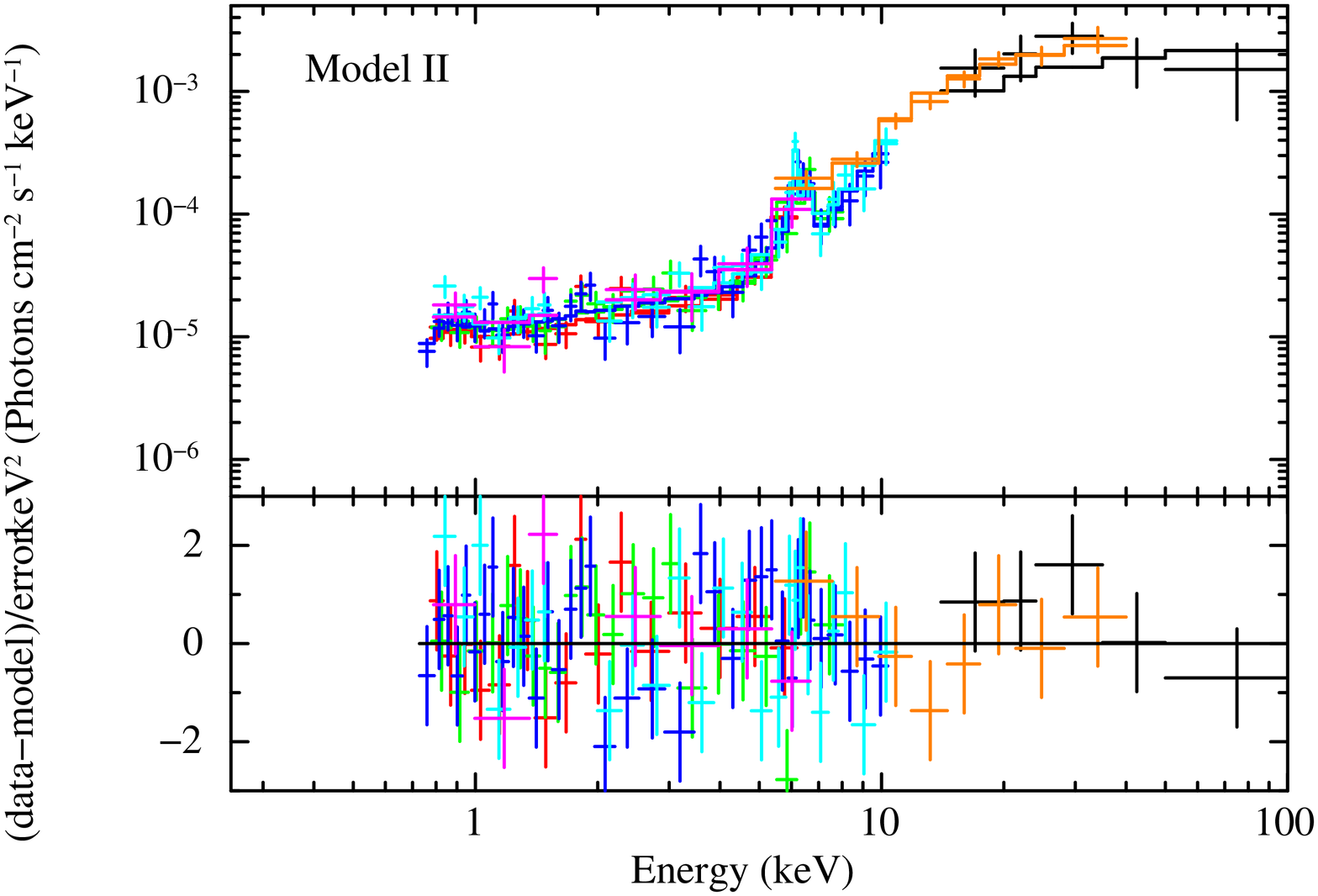}
\end{center}
\caption{Unfolded spectra for Model Ia, Model Ib, and Model II. 
The black, orange, light blue, magenta, green, blue, and red crosses are the
data of BAT, FPMs, FI-XISs, BI-XIS, EPIC-MOS, EPIC-pn, and ACIS,
respectively. The solid lines represent the best-fit models.}
\label{spectra}
\end{figure}

\begin{figure}
\begin{center}
\plotone{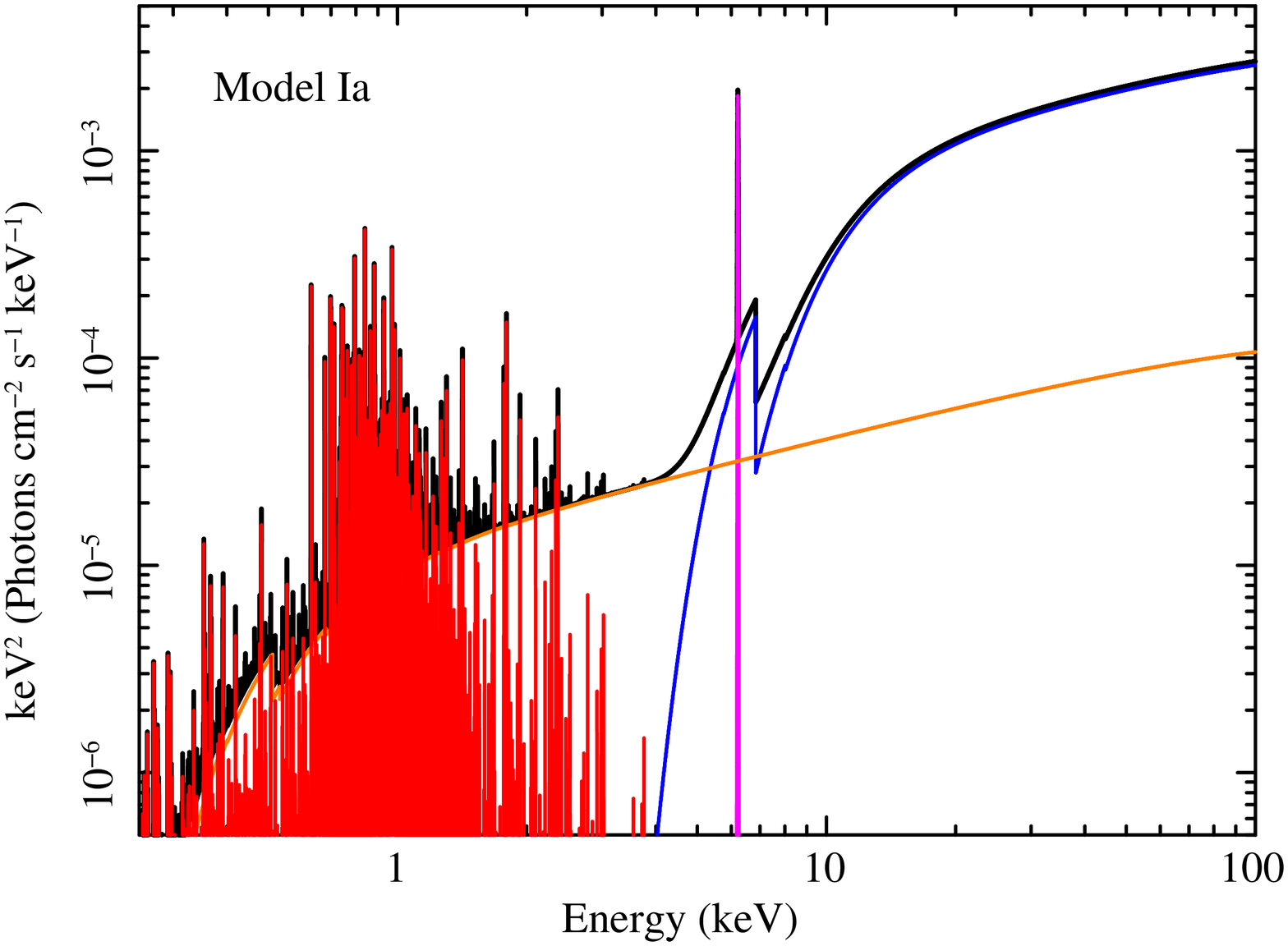}
\plotone{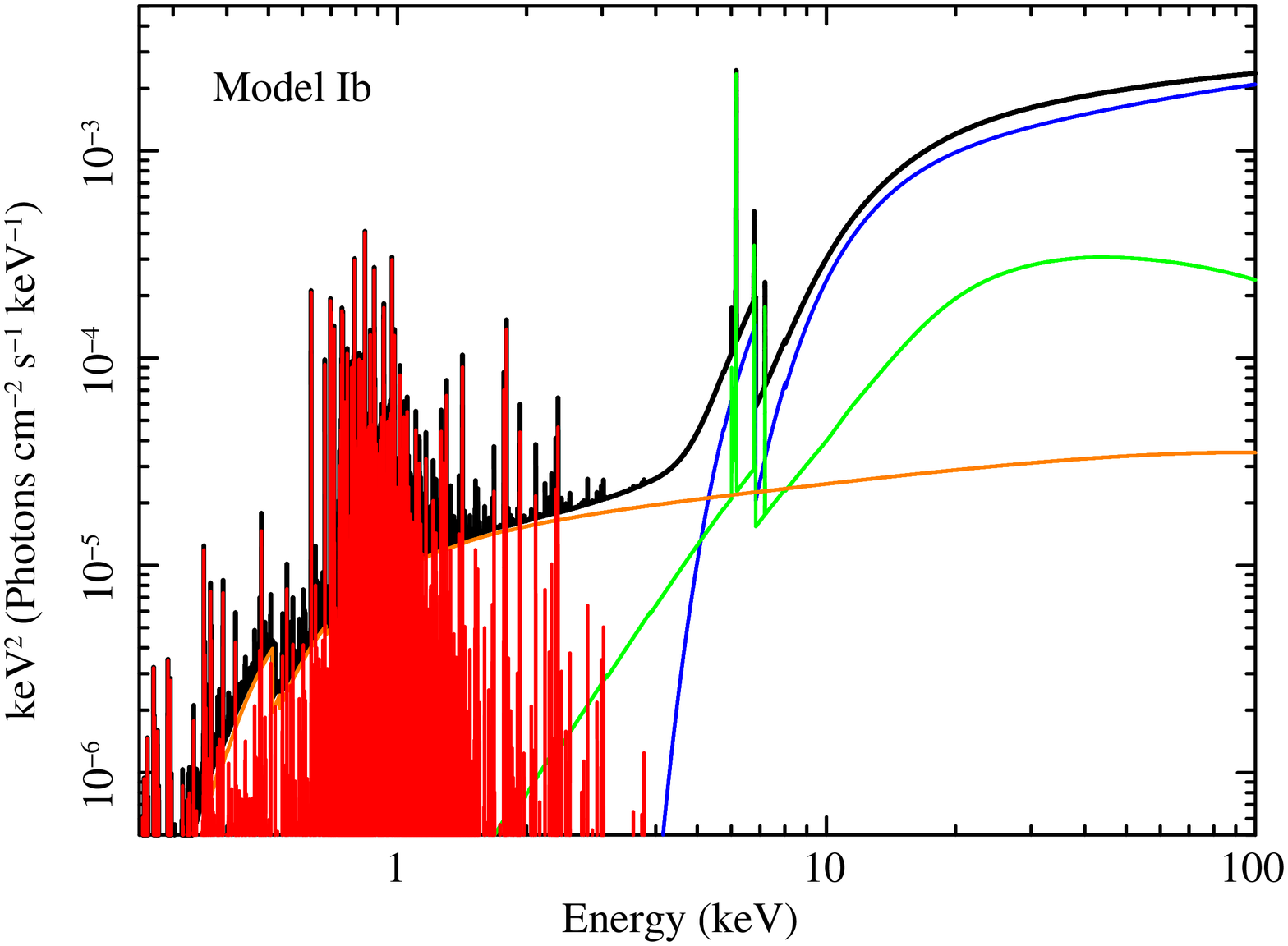}
\plotone{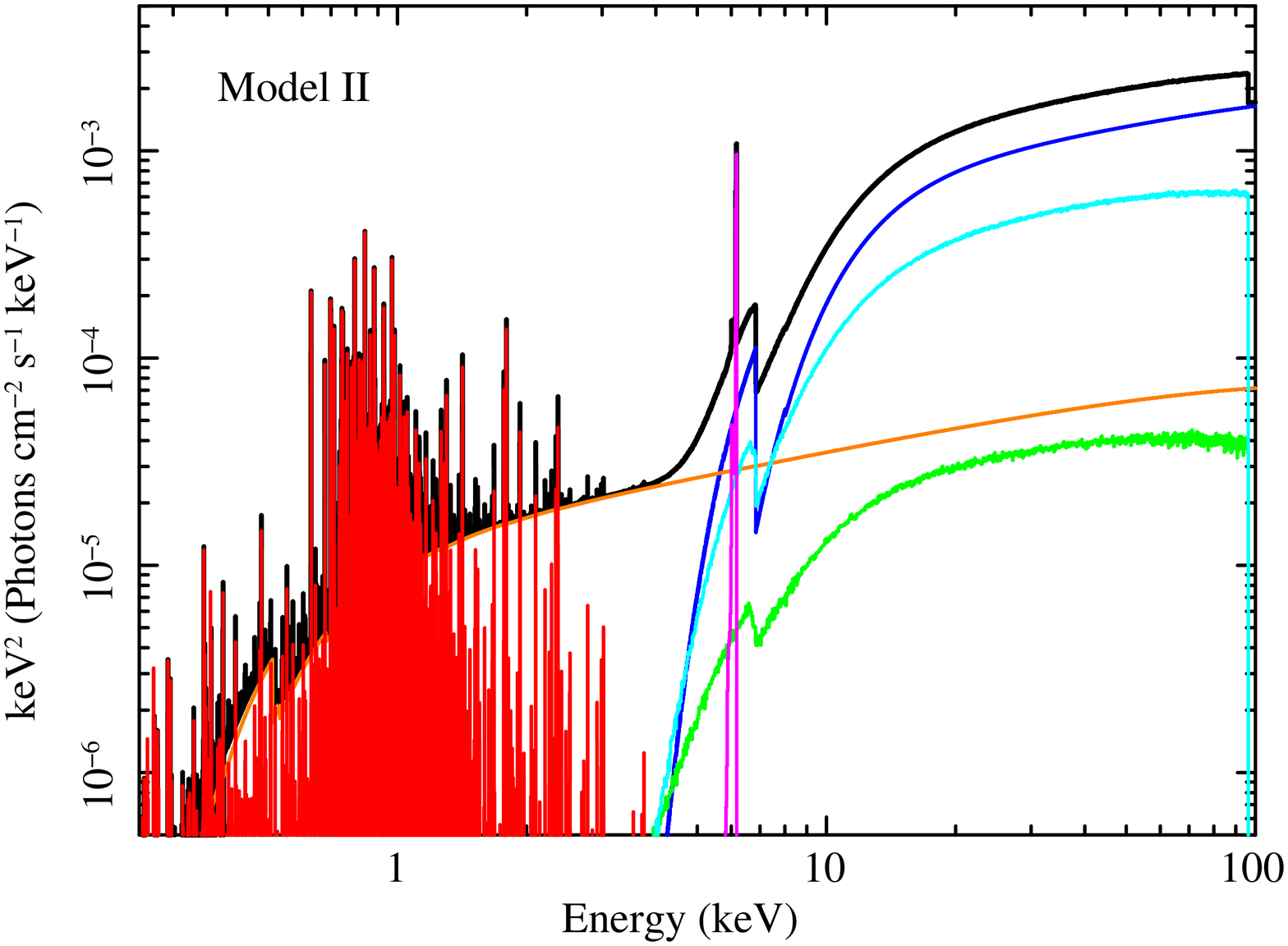}
\end{center}
\caption{Best-fit models for Model Ia, Model Ib, and Model II.
 For Model Ia, the black, blue, orange, magenta, and red lines represent
 the total, directly transmitted component, scattered component, iron-K$\alpha$ emission line,
 and thermal emission, respectively. 
 For Model Ib, the black, blue, orange, green, and red
 lines represent the total, directly transmitted component, scattered
 component, reflection component by \textbf{pexmon},
 and thermal emission, respectively.
 For Model II, the black, blue, orange, green, light blue, magenta, and red
 lines represent the total, directly transmitted component,
 scattered component, absorbed reflection component, unabsorbed
 reflection component, iron-K$\alpha$ emission line, and thermal emission, respectively.}
\label{eem}
\end{figure}

This model also well reproduces the combined spectra of UGC 5101 in the
0.7--100 keV band ($\chi^2$/d.o.f = 126.22/111).
Table~\ref{par2} lists the best-fit parameters, the observed fluxes and
intrinsic luminosities in the 2--10 keV and 10--50 keV bands
based on the \Swift/BAT spectrum, and the
iron-K line equivalent width with respect to the total continuum.
The unfolded spectra
and best-fit model are plotted in Figures~\ref{spectra} and \ref{eem},
respectively. The line-of-sight column density calculated from the
equation (3) in \citet{ike09} becomes $N_{\rm H}^\mathrm{LS} = 1.31^{+0.31}_{-0.36} \times 10^{24}$
cm$^{-2}$, confirming that the galaxy contains a Compton thick AGN. The
data prefer a large half-opening angle, $\theta_\mathrm{op} > 41$ degrees, with the
best-fit of 70 degrees (the maximum value in the table model).
With this geometry ($\theta_\mathrm{op} = 70$ 
and $\theta_\mathrm{inc} = 80$), the fraction of the reflection
component from the far-side torus absorbed (unabsorbed) by the near-side torus is 
$\approx 7$\% ($\approx 93$\%).
We find that the 10--50 keV intrinsic luminosity would be reduced 
by $\sim$6\% from our best estimate within the uncertainty in the torus
opening angle (41--70 degrees). This uncertainty does not affect our
conclusions.

\tabletypesize{\footnotesize}
\begin{deluxetable*}{cccc}
\tablecaption{Best fit parameters of analytical models  \label{par1}}
\tablewidth{0pt}
\tablehead{
 Note$^{*}$ & Parameter  &   Model Ia  &   Model Ib  
}
\startdata
(1) &  $ N_\mathrm{H}^\mathrm{host} \ [10^{21} \ \mathrm{cm^{-2}} ] $   & $1.06^{+0.80}_{-0.60}$  & $1.46^{+0.88}_{-0.66}$ \\
(2) &  $ N_\mathrm{H}^\mathrm{LS} \  [10^{24} \ \mathrm{cm^{-2} }]$   & $1.11^{+0.21}_{-0.27}$ &$1.26^{+0.33}_{-0.31}$ \\
(3) &  $\Gamma_\mathrm{AGN}$ &  $1.40 ^{+0.36}_{-0.30}$  & $1.61^{+0.34}_{-0.31}$\\
(4) & $A_\mathrm{AGN} \ [10^{-3} \ \mathrm{keV^{-1}\ cm^{-2} \ s^{-1}}]$   & $0.23^{+0.55}_{-0.16}$ &$1.0^{+3.3}_{-0.8}$ \\
(5) &  $R$ & -- & $-0.10^{+0.06}_{-0.15} $\\
(6) & $E_\mathrm{cent}\ [\mathrm{keV}] $   &  $6.46\pm0.04$ & --\\
(7) & $A_\mathrm{iron}\ [10^{-6} \ \mathrm{cm^{-2} \ s^{-1}}]$   &  $1.20^{+0.44}_{-0.43}$  & -- \\
(8) & $f_\mathrm{scat}\ [\%] $   &  $6^{+12}_{-4}$   & $1.4^{+4.5}_{-1.1}$  \\
(9)  &  $\Gamma_\mathrm{scat}$ &  $1.47^{+0.23}_{-0.21}$ & $1.74^{+0.27}_{-0.25}$ \\
(10) & $k \mathrm{T} \ [\mathrm{keV}] $   &  $0.90 \pm 0.12$  & $0.88^{+0.13}_{-0.12}$ \\  
(11) &  $ A_\mathrm{apec}\ [10^{-6} \ \mathrm{cm^{-5}}]$  &  $2.2^{+0.9}_{-0.8}$  & $2.0^{+0.9}_{-0.8}$\\
(12) & $N_\mathrm{FI}$  &  $1.06^{+0.14}_{-0.13}$  & $1.07 \pm 0.13$ \\
(13) &  $N_\mathrm{BI}$  &  $1.04^{+0.16}_{-0.21}$ & $1.05^{+0.15}_{-0.21}$ \\
(14) & $N_\mathrm{MOS}$   &  $0.92^{+0.19}_{-0.11}$  & $0.92^{+0.18}_{-0.11}$ \\
(15) &  $N_\mathrm{pn}$  & $0.93^{+0.19}_{-0.10}$  & $0.94^{+0.18}_{-0.10}$ \\
(16) &$N_\mathrm{ACIS}$  & 0.80 ($<0.96$)  & 0.80 ($<0.95$) \\
(17) &  $N_\mathrm{FPM}^\mathrm{time}$  & $1.69^{+0.45}_{-0.35}$ & $1.63^{+0.42}_{-0.32}$\\
(18) &  $F_{2-10} \ [\mathrm{erg \ cm^{-2} \ s^{-1}}]$  & $1.9 \times 10^{-13}$ & $1.9 \times 10^{-13}$ \\
(19) &  $F_{10-50} \ [\mathrm{erg \ cm^{-2} \ s^{-1}}]$  & $3.2 \times 10^{-12}$ & $3.2 \times 10^{-12}$ \\
(20) &  $L_{2-10} \ [\mathrm{erg \ s^{-1}}]$  & $ 5.1 \times 10^{42}$ & $ 1.6 \times 10^{43}$ \\
(21) &  $L_{10-50} \ [\mathrm{erg \ s^{-1}}]$  &  $1.2 \times 10^{43}$ & $ 2.9 \times 10^{43}$ \\
(22) &  $ \mathrm{EW} \ [\mathrm{eV}]$ & 361 (1433) & 202 (1433) \\
  &  $\chi^2/$dof   &  123.40 / 117   & 130.59 / 118 \\
\enddata 
\tablenotetext{*}{
(1)~The hydrogen column density for all AGN components.  
(2)~The hydrogen column density for the AGN transmitted component.  
(3)~The power-law photon index of the AGN transmitted component.
(4)~The power-law normalization of the AGN transmitted component at 1 keV.  
(5)~The relative reflection intensity. 
(6)~The central energy of the iron-K$\alpha$ emission line.  
(7)~The normalization of the iron-K$\alpha$ emission line.  
(8)~The scattering fraction.  
(9)~The power-law photon index of the scattered component.
(10)~The temperature of the thermal component.  
(11)~The normalization of the thermal component.  
(12)~The cross normalization of FI-XISs relative to BAT.  
(13)~The cross normalization of BI-XIS relative to BAT.  
(14)~The cross normalization of EPIC-MOSs relative to BAT.  
(15)~The cross normalization of EPIC-pn relative to BAT.  
(16)~The cross normalization of ACIS relative to BAT.  
(17)~The time variability normalization of FPMs relative to BAT.  
(18)~The observed flux in the 2--10 keV band.  
(19)~The observed flux in the 10--50 keV band.  
(20)~The de-absorbed AGN luminosity in the 2--10 keV band.  
(21)~The de-absorbed AGN luminosity in the 10--50 keV band.  
(22)~The equivalent width of the iron-K emission line with respect to
 the total continuum (the reflection continuum).} 
\end{deluxetable*}

\tabletypesize{\footnotesize}
\begin{deluxetable}{ccc}
\tablecaption{Best fit parameters of Model II \label{par2}}
\tablehead{
 Note$^{*}$  & Parameter  &   \\
 }
\startdata
(1) &  $ N_\mathrm{H}^\mathrm{host} \ [10^{21} \ \mathrm{cm^{-2}} ] $   &  $1.46^\dagger $ \\
(2) &  $ N_\mathrm{H}^\mathrm{Eq} \  [10^{24} \ \mathrm{cm^{-2} }]$   &  $1.32^{+0.32}_{-0.37}$ \\ 
(3) &  $\theta_\mathrm{op}\  [\mathrm{degree}]$ & 70 ($> 41$)  \\ 
(4) &  $\Gamma_\mathrm{AGN}$  &  $1.63 ^{+0.27}_{-0.13}$  \\
(5) & $A_\mathrm{AGN} \ [10^{-3} \ \mathrm{keV^{-1}\ cm^{-2} \ s^{-1}}]$  & $0.9^{+2.0}_{-0.5}$   \\
(6) & $f_\mathrm{scat}\ [\%] $   &  $1.5^{+1.5}_{-1.0}$  \\
(7) &  $\Gamma_\mathrm{scat}$ &  $1.58^{+0.13}_{-0.14}$ \\
(8) &  $k \mathrm{T} \ [\mathrm{keV}] $   &  $0.88^\dagger $   \\
(9) &  $ A_\mathrm{apec}\ [10^{-6} \ \mathrm{cm^{-5}}]$  &  $2.0^\dagger $  \\
(10) & $N_\mathrm{FI}$  &  $1.07^{+0.13}_{-0.13}$\\
(11) & $N_\mathrm{BI}$  &  $1.04^{+0.16}_{-0.22}$\\
(12) & $N_\mathrm{MOS}$   &  $0.92^{+0.21}_{-0.11}$\\
(13)  &  $N_\mathrm{pn}$  & $0.93^{+0.21}_{-0.10}$  \\
(14) &  $N_\mathrm{ACIS}$  & 0.80 ($< 1.0$)  \\
(15) &  $N_\mathrm{FPM}^\mathrm{time}$  &  $1.54^{+0.36}_{-0.29}$ \\
(16) &  $F_{2-10} \ [\mathrm{erg \ cm^{-2} \ s^{-1}}]$  & $2.0 \times 10^{-13}$ \\
(17) &  $F_{10-50} \ [\mathrm{erg \ cm^{-2} \ s^{-1}}]$  & $3.3 \times 10^{-12}$ \\
(18) &  $L_{2-10} \ [\mathrm{erg \ s^{-1}}]$  & $ 1.4 \times 10^{43}$ \\
(19) &  $L_{10-50} \ [\mathrm{erg \ s^{-1}}]$  &  $2.4 \times 10^{43}$ \\
(20) &  $ \mathrm{EW} \ [\mathrm{eV}]$ & 245 \\
  &  $\chi^2/$dof   &  126.22 / 111  \\
\enddata
\tablenotetext{*}{
(1)~The hydrogen column density for all AGN components.  
(2)~The hydrogen column density for the AGN transmitted component.  
(3)~The half opening angle of the torus.   
(4)~The power-law photon index of the AGN transmitted component.  
(5)~The power-law normalization of the AGN transmitted component at 1~keV.  
(6)~The scattering fraction.   
(7)~The power-law photon index of the scattered component.
(8)~The temperature of the thermal component.  
(9)~The normalization of the thermal component.  
(10)~The cross normalization of FI-XISs relative to BAT.  
(11)~The cross normalization of BI-XIS relative to BAT.  
(12)~The cross normalization of EPIC-MOSs relative to BAT.  
(13)~The cross normalization of EPIC-pn relative to BAT.  
(14)~The cross normalization of ACIS relative to BAT.  
(15)~The time variability normalization of FPMs relative to BAT.  
(16)~The observed flux in the 2--10 keV band.  
(17)~The observed flux in the 10--50 keV band.  
(18)~The de-absorbed AGN luminosity in the 2--10 keV band.  
(19)~The de-absorbed AGN luminosity in the 10--50 keV band.  
(20)~The equivalent width of the iron-K emission line
 with respect to the total continuum. }
\tablenotetext{$\dagger$}{
fixed at the same values as in Model Ib. 
} 
\end{deluxetable}

\section{Discussion}

We have presented the first high-quality broadband X-ray spectra of UGC
5101 in the 0.25--100 keV band, observed with \Swift, \NuSTAR, \Suzaku,
\XMM, and \Chandra. The combined data, whose total exposure except
\Swift/BAT is 185.5 ksec, have enabled us to best constrain the spectral
properties of this target. Besides the AGN contribution, we confirm the
presence of optically-thin thermal emission at low energies ($< 2$keV)
as previously reported by \citet{ima03}. The temperature ($\approx$0.88
keV) and luminosity ($\approx 1.1\times 10^{39}$ erg s$^{-1}$ in the
2--10 keV band) of the thin thermal component, which is spatially
resolved in the \Chandra \ image (\citealt{ima03}), is typical of ULIRGs
(e.g., \citealt{iwa11}).

As for the AGN contribution, we have shown that two analytical models
(Models Ia and Ib) and Monte-Carlo based numerical torus model (Model
II) can well reproduce the data in terms of $\chi^2$ test. Model Ia is a
simple, phenomenological model, while Model Ib is a more physically
motivated model including a reflection component from Compton-thick
matter. In Model II, a more realistic geometry of the torus is
considered. With our analysis, we firmly confirm the presence of a luminous
Compton-thick AGN in UGC~5101, whose transmitted component is detected
in the hard X-ray band above 10 keV.
The basic spectral parameters (photon index
and absorption) are consistent within their uncertainties
among the three models. We note, however, that if the 
Compton reflection component, required by the data, is not considered
in the model, a slightly flatter intrinsic photon index is derived
for the primary power-law component, i.e., 
$\Gamma_\mathrm{AGN} \approx$1.4 (Model~Ia) to be compared with  $\Gamma_\mathrm{AGN} \approx$1.6 
(Model~Ib and Model~II). 
More importantly, we find that if Compton scattering of the transmitted
component is ignored (Model~Ia), the intrinsic X-ray luminosity
is significantly underestimated compared with Model~II: a factor $\sim$2 in the 10--50 keV band
and a factor $\sim$3 in the 2--10 keV band (see Tables~4 and 5).
The Model~Ia result is very similar to the previous estimate by
\citet{ima03} ($\approx 6\times10^{42}$ erg s$^{-1}$), in which the
Compton scattering effects were also ignored.
By contrast, in Model~Ib, we apply the \textbf{cabs}
model to the transmitted component, assuming an extreme geometry where
the absorber is located only in the line of sight (see Section~3.1.2). In
this case, the luminosities are slightly overestimated compared with Model~II.
Hereafter, we refer to the luminosities obtained with Model~II unless
otherwise stated, because we consider it the most physically
self-consistent model among the three. 
We note that all luminosities listed in Table~4 and 5
are based on the
\Swift/BAT spectrum, because we are interested in time averaged values
when discussing the spectral energy distribution.  Even if we adopt the
luminosity obtained with \NuSTAR, which is higher than the \Swift/BAT one
by a factor of $1.54^{+0.36}_{-0.29}$, our main conclusions are
unaffected.

The fitting result with the torus model of \citet{ike09}, 
$\theta_\mathrm{op}> 41$ degrees, suggests that the 
covering fraction of Compton-thick parts of the torus
is only moderate.
This result is consistent with the best-fit scattering fraction,
$f_\mathrm{scat} \sim 1.5 \%$, which implies that UGC 5101 is not a
``low scattering fraction AGN'' with $f_\mathrm{scat} < 0.5 \%$, putative
population of AGNs with small torus opening angles discovered by hard
X-ray surveys (\citealt{ueda07}).
The scattering fraction of UGC 5101 is similar to, and even larger
than, the median values of an [O~III]$\lambda 5007$ selected sample
($f_\mathrm{scat} \sim 1.5\%$; \citealt{nog10}) and a hard X-ray (14--195 keV) selected 
sample ($f_\mathrm{scat} \sim 1\%$; \citealt{kaw16}), respectively.
We note, however, that given the large error in the scattering
fraction and possible contamination by emission from high mass X-ray
binaries in the host galaxy, we cannot make strong
arguments based on the $f_\mathrm{scat}$ value.

These results imply that the nucleus in UGC 5101 cannot be
fully surrounded by Compton-thick matter.
Nevertheless, there could be
the possibility that a lower column-density 
(Compton-thin) absorber is present in the
torus-hole regions that attenuates the nuclear 
ionizing UV emission and causing
the under-developed NLR in UGC 5101 (Section 1).
By our analysis, we indeed find that 
a weak intrinsic X-ray absorption ($N_\mathrm{H}^\mathrm{host} \sim 10^{21}$ cm$^{-2}$,
\textbf{zphabs1} in Section~3), in addition to the Compton-thick absorber,
is required
when the data below 0.7 keV, which are sensitive to determine
such an absorption, are included in the fit (i.e., Models~Ia and Ib).
By assuming a Galactic dust-to-gas ratio, this column density
corresponds to an optical extinction of $A_V \sim 0.5$ mag
(\citealt{guv09}). This is sufficient to block a large fraction of
ionizing UV lights from the central engine of the AGN, although we
cannot discard the possibility that the absorption could be associated
to interstellar medium of the host galaxy on a larger scale.

\begin{figure}
\begin{center}
\plotone{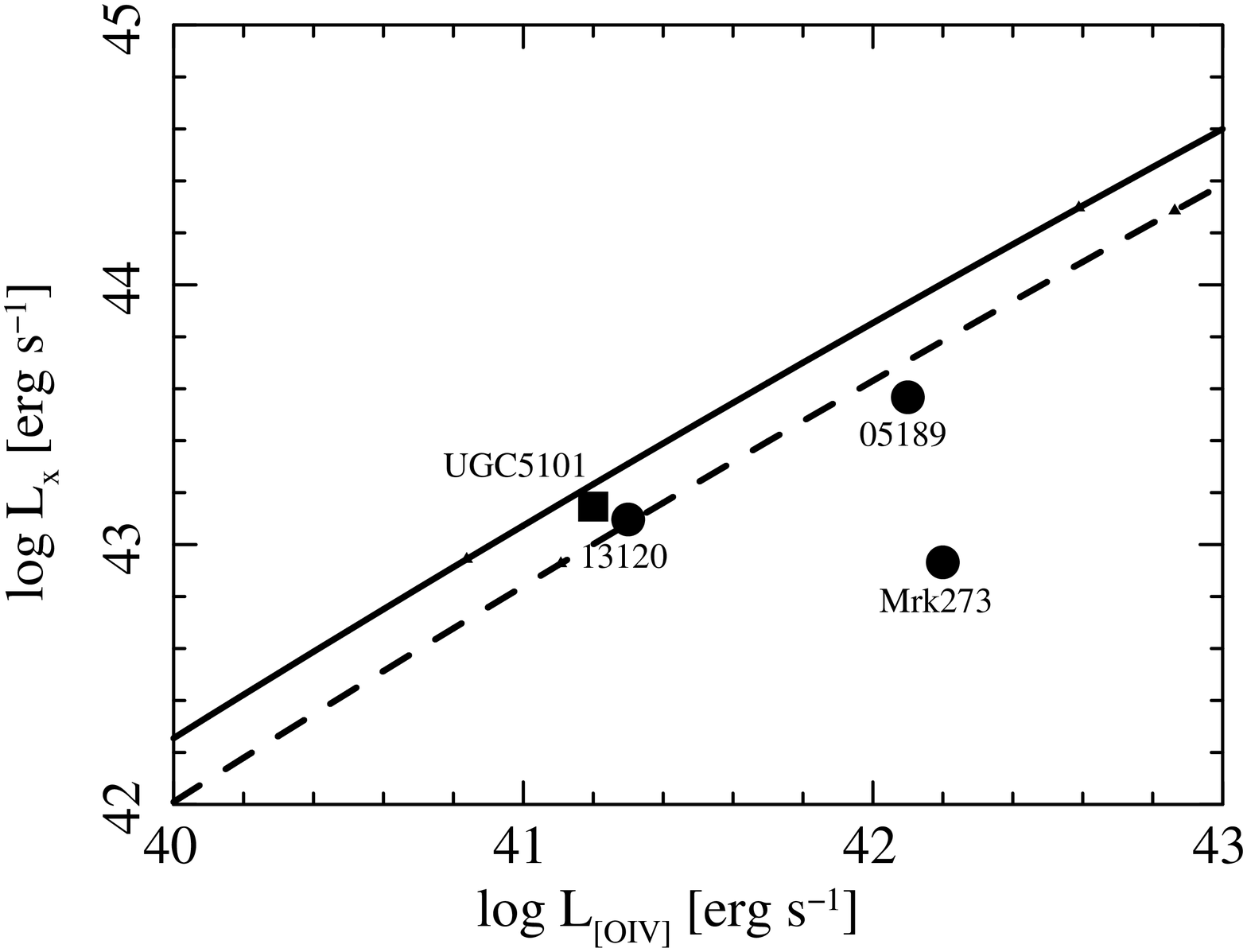}
\end{center}
\caption{Comparison between the absorption-corrected 2--10 keV AGN luminosity and the [O IV] 26
$\mu$m luminosity for UGC 5101 (square) and typical ULIRGs observed by
\citet{teng15} (IRAS 05189--2524, IRAS 13120--5453, and Mrk 273, circles). The solid and dashed lines represent the expected
relations based on \citet{rig09} for Seyfert 1 and Compton-thin Seyfert 2
galaxies,respectively.}
\label{oiv}
\end{figure}

To further investigate 
the nuclear geometry and intrinsic spectral energy distribution, 
we compare the luminosity of an emission line from the NLR and
the AGN X-ray luminosity.
Figure \ref{oiv} plots the correlation between the 
[O IV] 26 $\mu$m luminosity and the absorption-corrected 2--10 keV AGN luminosity for UGC 5101
(square). 
Assuming that the ratio between the [O IV] 26 $\mu$m and
far-infrared (FIR) luminosity in pure starburst galaxies is typically $L_{\rm
[O\ IV]}/L_{\rm FIR} \sim 10^{-5}$ (e.g., see Figure 22 of
\citealt{vei09}),
we estimate that starburst contamination in the observed [O IV] 26 $\mu$m luminosity is very small, 
$\sim (1-f_{\rm AGN}) \times 10\%$,
where $f_{\rm AGN}$ is the
fractional AGN contribution to the total bolometric luminosity (see below).
The flux ratio between [Ne V] 14.3 $\mu$m and [O IV]
26 $\mu$m lines also supports that the [O IV] luminosity is dominated by
the AGN component (\citealt{arm07}). For comparison, we also plot the data of other
ULIRGs, 
IRAS 05189--2524, IRAS 13120--5453, and Mrk 273, taken from 
the sample observed by \citet{teng15} by excluding Superantenna (IRAS F19254--7245),
a candidate of a heavily Compton thick
AGN ($N_\mathrm{H}^\mathrm{LS} > 10^{25}$ cm$^{-2}$), and Mrk 231, a probable
outlier object (broad absorption line QSO). The curves represent the
expected relations for Seyfert 1 (solid) and Compton-thin Seyfert 2
galaxies (dashed) based on the average L$_\mathrm{bol}$/L$_\mathrm{[O\ IV]}$ ratio obtained by
\citet{rig09} with the 2--10 keV to bolometric luminosity conversion factor
given by \citet{mar04}.

As noticed from Figure~\ref{oiv}, the [O IV] to X-ray luminosity ratio
in UGC 5101 is similar to those of
normal Seyfert galaxies. 
This is also the case for the two ULIRGs except Mrk 273, 
demonstrating that a significant
portion of ULIRGs in the current \NuSTAR\ sample are
not really ``X-ray faint'' 
with respect to the [O IV] 26 $\mu$m flux.
Since the luminosity of [O IV] 26 $\mu$m is determined by that of
ionizing UV photons leaked to the NLR out of surrounding matter, 
it should be roughly proportional to the product of ( 1 $-$ covering fraction ) and
the intrinsic AGN luminosity in the UV band.
Thus, there could be two explanations for this observational fact. 
The first one is that the AGN in UGC 5101 is not different from those in
normal Seyfert galaxies in terms of the covering fraction of surrounding
matter and
UV to X-ray spectral energy distribution (SED). The second one is that,
in UGC 5101, \textit{both} [O IV] 26 $\mu$m and X-ray luminosities are
smaller with respect to the bolometric AGN luminosity than
in normal Seyfert galaxies.

We find difficulties in the first explanation, however. 
The AGN bolometric luminosity converted from the 2--10 keV luminosity
with the \citet{mar04} relation (i.e., that for normal AGNs) turns out
to be $\approx 2.6 \times10^{44}$ erg s$^{-1}$ (with a bolometric
correction factor of 19), corresponding to only $\approx$7\% of the
total bolometric luminosity. This fraction is smaller than those
estimated from the {\it Spitzer} spectra with diagnostics not using the
NLR lines, $f_{\rm AGN}=$52\%--80\% \citep{arm07} (see their Table~7) or
$f_{\rm AGN}=$14\%--55\% \citep{vei09} (see their Table~12). This
implies that the SED of the AGN in UGC 5101 is X-ray under-luminous
relative to the UV luminosity, which dominates the bolometric one, as
compared with normal AGNs.
Also, as mentioned in Section~1, the [O IV] 26 $\mu$m luminosity 
seems to underestimate the true bolometric AGN luminosity, 
implying that the total covering fraction of 
Compton-thick and Compton-thin matter is larger than those of normal
AGNs.

\begin{figure}
\begin{center}
\plotone{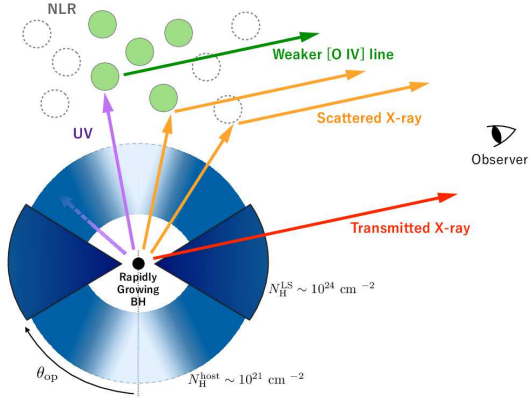}
\end{center}
\caption{Geometry of our proposed scenario described in
Section~4. The Compton-thick torus has only a moderate covering
fraction. The torus-hole region is partially filled with gas and dust of
$N_\mathrm{H}^\mathrm{host} \sim 10^{21}$~cm$^{-2}$. Thus, a significant
fraction of ionizing UV photons from the central AGN are blocked before
reaching the NLR, while X-rays are not blocked thanks to their strong
penetrating power.
} \label{torus}
\end{figure}

We therefore consider that the second explanation would be a likely scenario
for UGC 5101. Our picture, illustrated in Figure~\ref{torus}, 
is summarized as follows, even though there are
large uncertainties (by a factor of a few) in the numbers.
The surrounding structure consists of two parts, a Compton-thick part ($N_{\rm
H} \sim 10^{24}$ cm$^{-2}$) located close to the equatorial plane and
a Compton-thin part ($N_{\rm H} \sim 10^{21}$ cm$^{-2}$) in the torus hole
region, although they may not be discrete in reality. 
The covering fraction of Compton-thick matter is small, 
which is derived from the hard X-ray
spectra. The {\it total} covering fraction of Compton-thick and
Compton-thin matter that blocks UV lights is large, which accounts for the low
[O IV] 26 $\mu$m luminosity relative to the AGN bolometric luminosity.
The intrinsic bolometric luminosity of the AGN is truly $L_\mathrm{bol}
\sim 2\times10^{45}$ erg s$^{-1}$, corresponding to $\sim$50\% of the
total luminosity, as inferred from the IR spectra. 
Because the
intrinsic AGN luminosity in the 2--10 keV band we determine is $L_{\rm
X} \approx 1.4\times10^{43}$ erg s$^{-1}$, the 2--10 keV to bolometric
luminosity correction factor of this AGN is estimated to be $ L_{\rm
bol} / L_{\rm X} \sim$140.  This value is relatively large
(i.e., the UV to X-ray SED is steep) compared with
those of typical Seyferts with similar X-ray luminosities \citep{mar04},
but can be expected at high Eddington rates \citep{vas07}
i.e., in rapidly growing SMBHs. Thus, because both the
[O IV] 26 $\mu$m and X-ray luminosities are suppressed, their ratio
remains almost the same as those of normal Seyferts. 
Nevertheless, we argue that AGNs
in many ULIRGs do not look extraordinary (i.e., extremely X-ray faint),
as suggested by recent works (e.g., \citealt{teng15}), compared with
normal Seyferts.
Since the strong starburst activity in the nuclear region may increase
both the total covering fraction of surrounding matter and the mass accretion
rate, it may not be a coincidence that a ULIRG shows a standard value of
the [O IV] 26 $\mu$m to X-ray luminosity ratio.

To summarize, we have revealed the complex structure of the
circumnuclear region of UGC 5101; in short, the
covering fraction of the Compton-thick torus is small, while the total
covering fraction of Compton-thick and Compton-thin matter is large,
making the nucleus ``unburied'' in X-rays but ``buried'' in UV. This
implies that, in a more realistic geometry, there is a strong contrast
in the column-density distribution as a function of inclination angle,
that is, matter is highly concentrated toward the equatorial plane. We
point out that the nearby Compton-thick AGN NGC 4945 may have a similar
circumnuclear structure to that of UGC 5101. In NGC 4945, X-ray spectral
and variability properties suggest a small covering fraction of the
Compton-thick torus (e.g., \citealt{ito08}; \citealt{mari12}),
whereas, from observations of NLR lines (e.g.,
\citealt{per11}), it is inferred that the AGN is surrounded by matter
with a large covering fraction. To answer the question if this is a
general property of obscuring material in Compton thick AGNs, it is
important to perform sensitive broadband X-ray observations for a larger
sample of heavily obscured AGNs.

\section{Conclusion}

Using all available data taken with \Swift, \NuSTAR, \Suzaku, \XMM,
and \Chandra, we have presented the best quality broadband X-ray spectra
of UGC 5101 covering the 0.25--100 keV band. The spectra are well
reproduced by conventional analytic models or a numerical torus model
for an obscured AGN. 
The main conclusions are summarized below.

\begin{itemize}

\item We detect a Compton-thick AGN with a line-of-sight column
      density of $N_\mathrm{H}^\mathrm{LS} \approx 1.3\times 10^{24}$ cm$^{-2}$. The
      intrinsic 2--10 keV luminosity is determined to be
      $\approx 1.4\times10^{43}$ erg s$^{-1}$, which is $\approx$2.5 times
      larger than the previous estimate using only data below 10 keV.

\item A weak absorption of $N_\mathrm{H}^\mathrm{host} \sim 10^{21}$
  cm$^{-2}$ to the soft X-ray scattered component is required.

\item A spectral fit with a smooth torus model favors a large ($>41$
  degrees) half opening angle of the torus. This suggests that the
  covering fraction of a Compton-thick part is moderate.

\item The ratio between the [O IV] 26 $\mu$m line and 2--10 keV
      luminosities of UGC 5101 and other typical ULIRGs is found to be
      similar to those of normal Seyfert galaxies. Thus, a significant
      portion of ULIRGs are not really ``X-ray faint'' with respect to
      the [O IV] 26 $\mu$m flux.

\item To reconcile these results with the large AGN contribution
  ($\sim$50\%) to the total bolometric luminosity inferred by the IR
  spectra, a possible scenario is that (1) the AGN in
  UGC 5101 is surrounded not only by
  Compton-thick matter located close to the equatorial plane 
  but also by Compton-thin matter with
  $N_\mathrm{H}^\mathrm{host} \sim 10^{21}$ cm$^{-2}$ in the torus-hole
  region, and (2) it has a steep UV to X-ray SED, which is expected at a
  high Eddington rate. These two effects work to somewhat (but not
  extraordinarily) reduce the [O IV] 26 $\mu$m and X-ray luminosities,
  respectively, by keeping their ratio almost the same as in normal
  Seyfert galaxies.

\item The circumnuclear structure of UGC 5101 may be similar to that of NGC
      4945, where the covering fraction of a
      Compton-thick part is small but the total covering fraction of
      Compton-thick and Compton-thin parts is large.

\end{itemize}

\bigskip

This research has made use of the \NuSTAR \  Data Analysis Software
(\textsc{NuSTARDAS}) jointly developed by the ASI Science Data Center
(ASDC, Italy and the California Institute of Technology (Caltech,
USA). Part of this work was financially supported by the Grant-in-Aid for
Scientific Research 26400228 (Y.U.), 15K05030 (M.I.) and 15H02070 (Y.T.)
from the Japan Society for the
Promotion of Science (JSPS), and by the CONICYT-Chile grants
``EMBIGGEN" Anillo ACT1101 (C.R.), FONDECYT 1141218 (C.R.), and Basal-CATA
PFB--06/2007 (C.R.).
C.R. acknowledges support from the China-CONICYT fund.

\end{document}